# Bounded Local Generator Classes for Deterministic State Evolution

*Banach-space formulation, temporal locality, and measured persistent-state execution*

**R. Jay Martin**
Independent Researcher
Lake Arrowhead, CA 92352
ray@opalone.ai

*Abstract*

Persistent memory must retain history without revisiting it at every transition. Bounded Local Generator Classes (BLGC) formalize carried numerical state, selected context and ordered local updates. The $\ell^\infty$ formulation permits countably many uniformly bounded coordinates without square summability. Projected maps preserve admissibility, Lipschitz estimates control finite-horizon sensitivity, and bounded acquisition and evaluation give a cardinality-independent transition-work contract.

OPAL supplies a working persistent-state engine and controlled production measurements. After six successful million-node workloads, 18 matched fresh Release processes compared original and incremental receipt indexing at 10,000, 100,000 and 1,000,000 nodes. Degree, record shape, state dimension, active-target count and event schedule were fixed. Repaired resident-update P50 remained 0.083-0.084 microseconds; subsequent receipt-event P50 ranged from 4.458 to 6.459 microseconds across all scales. The original receipt path took 177.700-182.297 milliseconds at one million nodes.

**Reliability was tested alongside performance.** Baseline and repaired engines produced byte-identical checkpoints, including continuation after fresh-process recovery. Replay, conflict, rejection, concurrent-event and allocation-failure checks passed; all 18 matched scaling processes passed their correctness gates. The repair preserves the state law and durable receipt format. These results connect the mathematical construction to a traced production transition with verified state continuity and measured scaling. Capacity-growth costs and formal scope are reported separately.

# Reading guide and notation

The paper connects temporal organization, a precise transition class, and production evidence. Its central question is whether carried state and selected context support bounded incremental work as retained memory grows.

**Sections I-III** present the architecture, state space and guarantees. **IV.2** summarizes operational reliability; **IV.11-IV.17** report the production mapping, controlled studies and receipt-index repair. **IV.18** consolidates scope. **Section V** draws the conclusions. Appendices A and B preserve the historical figure and mathematical continuity.

| Symbol | Meaning in the formal model |
|---|---|
| V, i | Countable coordinate index set; target index |
| d | Fixed numerical dimension of one local state |
| $\Psi$, $s_i$ | Global configuration; its coordinate at i |
| X, S | Bounded-sequence Banach space; admissible unit ball |
| t, $\pi(t)$ | Transition index; scheduled target |
| $A_i(t)$, D | Active dependency set; uniform cardinality bound |
| $f_{i,t}$, $L_{i,t}$ | Local functional; its Lipschitz bound |
| $\eta$, $\Pi$ | Positive step parameter; Euclidean unit-ball projection |
| $G_{i,t}$, g(t) | One coordinate update; ordered composition |
| M | Node count of a finite instance only |

The discrete transition index t is not a wall-clock timestamp. Elapsed time and validated external conditions may parameterize a particular functional, but this manuscript does not prescribe the engine's private realization. The same symbol is never used to identify a numerical state with an entire stored record.

# I. Systems-native formulation

## I.1 The scaling question and contribution

Persistent memory must retain information, accept changes and supply state for the next transition. Reconstructing that state from the full retained collection makes each change more expensive as memory grows. The architectural goal is to control repeated work while retaining a useful past.

BLGC organizes this work temporally: a target carries its state forward, a local functional applies selected validated context, and an ordered commit supplies the next state. Earlier information can influence the present through carried state without replaying every prior event.

The contribution has three parts: an $\ell^\infty$ formulation accommodating countably many uniformly bounded local states; projected-coordinate, finite-horizon sensitivity and transition-work results; and a production study of state continuity and scaling. The study traces a resident transition, removes repeated receipt-index reconstruction, and verifies matching canonical outcomes before and after the repair.

Theorem 3 states the cost contract: bounded acquisition and evaluation yield bounded transition work. IV.1 maps the implementation obligations; IV.15-IV.17 test the repaired production boundary. The contribution is this formulation, temporal interpretation and operational evidence. Priority over all local-update systems is not claimed.

Retaining information need not require revisiting it at every step. The engineering contracts test continuity and execution; the usefulness of carried information for a particular application remains an empirical question.

## I.2 Graph-indexed states and separate information domains

Let $\Gamma = (V, E)$ represent a collection of identities and their stored relationships, with V countable. A finite instance has $M = |V|$ nodes. For the conditional numerical model, each $i \in V$ carries a vector $s_i \in \mathbb{R}^d$, where d is fixed, and

$$\|s_i\|_2 \le 1.$$

The configuration is $\Psi = (s_i)_{i \in V}$. Its admissible domain is the product of the local Euclidean unit balls. Section II identifies this set with the closed unit ball of a bounded-sequence Banach space.

This definition separates numerical evolution from the other information required by a memory system. Identity, content, provenance, timestamps, policy, history, and relationships are not silently packed into $\mathbb{R}^d$. Variable-length text and an indefinitely retained event ledger, for example, do not become bounded Euclidean coordinates merely because they belong to the same stored object.

A configuration collects local coordinate states. Its notation describes the whole system, while an implementation can update one stored coordinate without copying the others.

Local Euclidean balls are compact. The infinite product considered with the supremum norm is generally not compact. Uniform coordinate boundedness, global compactness, regularity of the update, and efficient evaluation are different properties. Separating them permits the proof to use exactly the property needed at each step.

Background on local interaction and state processes is supplied by Bratteli and Robinson, Lieb and Robinson, and Liggett [2, 5, 6]. Their settings differ from the temporal interpretation here; no result about OPAL is imported from those works by analogy.

## I.3 Temporal locality and active dependency support

Temporal locality concerns the information available at successive state transitions. The target state supplies continuity with its preceding state. Selected contextual states supply additional inputs. Let $A_i(t) \subseteq V$ denote the indices inspected by the local functional at transition t, and assume

$$|A_i(t)| \le D,$$

where D is independent of total retained cardinality M. The target $s_i$ is available separately and is counted once in the dependency union $A_i(t) \cup \{i\}$.

The dependency support separates carried target state from acquired context. The graph may grow while each application uses the target and at most D additional vectors.

A realization must establish both the cardinality bound and the procedure for acquiring those inputs. Replaying an unbounded history or scanning a growing adjacency list before presenting a small tuple would move the scaling cost into acquisition. Theorem 3 therefore requires bounded acquisition explicitly; it does not derive that property from the name temporal locality.

A radius-r graph neighborhood is an optional specialization. It requires an independent degree or neighborhood-cardinality bound. Temporal locality itself is defined here by continuity and selected transition context, rather than graph distance.

## I.4 Relationships, traversal, and evolution

Stored relationships describe available links; traversal follows links for a query; temporal evolution changes state between transitions. Their operational connection must be specified.

A transition need not traverse the graph simply because its target has relationships. Conversely, a query may inspect many candidates even when a particular numerical transition has bounded support. A hop limit bounds path length; it does not automatically bound the number of adjacency entries examined to choose each hop.

A global geometry or Flow summary carries its acquisition cost even when its output is small. Caching, incremental maintenance and scheduled recomputation are possible implementations; the formal proof does not choose among them.

The graph index remains useful for naming states and describing optional dependencies. It does not replace temporal state-space locality with a static spatial neighborhood. Similarly, a retrieval result is not automatically an evolution event. Whether access reinforces state is an explicit operational contract.

Section IV reports indexed lookup, traversal, persistence, global Flow processing and temporal tests as distinct operations with distinct timing boundaries.

## I.5 Local generators: definition and explanation

For an admissible configuration $\Psi$, a target i, and fixed context at transition t, define

$$T_{i,t}(\Psi) = \Pi(s_i + \eta\, f_{i,t}(\Psi|A_i(t))),\ \eta > 0.$$

The global map $G_{i,t}$ replaces coordinate i by $T_{i,t}(\Psi)$, leaving every coordinate $k \neq i$ unchanged. Here $\Pi$ is Euclidean projection onto the closed unit ball. The functional $f_{i,t}$ is assumed Lipschitz on its admissible input domain; its evaluation cost is an additional assumption, stated in Section II.

There are two different supports. Modification support is the single target coordinate. Dependency support is contained in $A_i(t) \cup \{i\}$. The latter describes what may influence the result, whereas the former describes what the transition changes.

This is a nonlinear coordinate map. It need not be a bounded linear operator, and it is not described as rank-one in the linear-algebraic sense. Projection and a nonlinear functional generally invalidate that terminology. Finite modification support is the property required here.

Projection supplies an admissible result even if the unprojected candidate lies outside the ball. Lipschitz continuity controls sensitivity to input perturbations. Neither property alone specifies the semantic meaning of a coordinate or proves that the functional is cheap to compute.

Local updates can produce global change: later transitions may depend on states altered earlier. Bounded modification support limits one step, not accumulated influence throughout the system.

## I.6 Deterministic evolution and temporal order

Let $\pi : \mathbb{N} \to V$ select the target at each transition index. With a specified sequence of contexts and active sets, evolution is

$$\Psi_{t+1} = G_{\pi(t),t}(\Psi_t).$$

Define the finite composition

$$g(t) = G_{\pi(t-1),t-1} \circ \cdots \circ G_{\pi(0),0}, \; g(0) = \mathrm{Id}.$$

Then $\Psi_t = g(t)(\Psi_0)$. The identity convention makes the zero-transition case explicit and preserves the original paper's composition-based explanation.

Determinism means that the initial configuration, controller state, ordered inputs, schedule, local functionals, and tie-breaking rules determine one trajectory. Equal initial numerical states alone are insufficient if the event order or external conditions differ. A recorded logical order is particularly important when multiple callers can submit operations.

Asynchronous execution need not advance every node in a global sweep. It must preserve dependencies and commit order, or prove that permitted reorderings are equivalent. Single-coordinate writes alone do not prevent overlapping read/write conflicts.

Deterministic mathematical maps also do not automatically guarantee bitwise portability across floating-point implementations. Numerical libraries, reduction order, compiler choices, and exceptional-value handling require their own contract. The engineering tests in Section IV provide scoped replay evidence; they are not an unrestricted claim about all hardware and schedules.

## I.7 The living-topology controller

A temporal system may alter which information is relevant to a later update. The active set can therefore change between transitions. The useful distinction is between continuous numerical state maps within a specified context and a discrete living-topology or context-selection process that chooses that context.

For a fixed active-set and schedule realization, the Lipschitz estimates below compare trajectories using the same maps in the same order. If a selector depends on evolving usage, state, or policy, two nearby trajectories may choose different active sets. Deterministic selection guarantees a reproducible choice for a given input; it does not guarantee continuity of that choice.

For example, a deterministic ranking rule can switch selected inputs when two scores cross. A fixed tie rule resolves equality, but nearby unequal scores on opposite sides of the boundary can still choose different inputs. A continuity proof for the composite selector-and-update process needs additional hypotheses. The present finite-horizon theorem does not supply them implicitly.

The temporal interpretation is preserved by treating context changes explicitly. A new policy, new topology, different dependency membership, revised input generation, or changed numerical dimension can begin a new regime. A proof applying inside the old regime must be reconsidered at that boundary.

## I.8 The contribution as an implementation contract

The construction connects five obligations: a bounded numerical domain, bounded dependency support, regular local maps, a uniform evaluation procedure, and stated physical-access conditions. Together they describe when retained cardinality can increase without increasing the work of one specified transition.

The first three obligations support state preservation and a finite-horizon sensitivity bound. The fourth supports the algorithmic-work result. The fifth is needed to relate that result to physical latency. Keeping these obligations distinct makes the construction usable: an implementation can identify the relevant state, trace its dependencies, and account for acquisition and evaluation before claiming the bound.

The temporal interpretation gives this contract its architectural purpose. A carried state is available to the next transition; selected context updates it; ordered composition allows earlier changes to influence later ones. A realization must also determine what this state preserves about its past. Fixed dimension provides a cost parameter, not a theorem of semantic sufficiency.

Sections II and III establish the conditional mathematics. IV.1 turns its requirements into an explicit correspondence register for OPAL. The engineering results then show tested continuity, persistence, and access behavior at their actual boundaries. IV.18 gathers the outstanding research obligations so that they can be evaluated together.

# II. Banach-space representation

## II.1 Bounded sequences and admissible configurations

Let V be nonempty and countable, and let $d \geq 1$. Define

$$X = \ell^\infty(V; \mathbb{R}^d),\ \|\Psi\|_\infty = \sup_{i\in V} \|s_i\|_2.$$

The space X consists of uniformly bounded sequences of Euclidean vectors. It is complete. Let

$$S = \{ \Psi \in X : \|\Psi\|_\infty \leq 1 \}.$$

Thus S is the closed unit ball of X, equivalently the configurations satisfying $\|s_i\|_2 \leq 1$ at every index. It is closed, bounded, convex, and complete in the induced metric.

The supremum norm measures the largest local magnitude, rather than summing energy over all nodes. For an infinite V, a constant nonzero bounded configuration belongs to X but need not belong to $\ell^2(V; \mathbb{R}^d)$. This is why the Banach representation fits uniform node-local bounds without imposing an additional square-summability requirement.

No supremum-norm compactness is asserted. Choose a fixed unit vector and place it at one different index in each member of a sequence, setting all other coordinates to zero. Distances between distinct members are one, so the sequence has no convergent subsequence in that norm.

For a finite restriction $V_M \subseteq V$ with $|V_M| = M$, the coordinate space $X_M = (\mathbb{R}^d)^{V_M}$ has dimension dM. This finite-dimensional identity is not assigned to the infinite space. A restricted autonomous evolution also needs defined active sets or boundary inputs; deleting coordinates alone does not define their replacement dynamics. Functional-analysis background is given in [7].

## II.2 Coordinatewise Euclidean projection

Let $B_1(0)$ be the closed Euclidean unit ball in $\mathbb{R}^d$. Its metric projection is

$$\Pi(x) = x / \max\{1, \|x\|_2\}.$$

Euclidean projection onto a closed convex set is nonexpansive:

$$\|\Pi(x) - \Pi(y)\|_2 \leq \|x - y\|_2.$$

Define the pointwise map $P : X \to S$ by $(P\Psi)_i = \Pi(s_i)$. Taking the supremum of the coordinate inequalities gives

$$\|P\Psi - P\Phi\|_\infty \le \|\Psi - \Phi\|_\infty.$$

This is a statement about the particular pointwise projection just defined. It is not a claim that arbitrary metric projections in every Banach space are nonexpansive. The Euclidean geometry of each local coordinate is doing the required work.

The proof of admissibility is equally direct. Every projected coordinate has norm at most one, so $P\Psi$ belongs to S. For a generator that changes only one coordinate, projection need only be applied at that coordinate; unchanged coordinates retain their previous admissible values.

Projection supplies confinement to the admissible domain; the separate Lipschitz analysis supplies finite-horizon sensitivity. Nonlinear optimization background is provided in [1].

## II.3 Active inputs, temporal context, and fixed regimes

For each target i and transition t, choose an active set $A_i(t) \subseteq V$ with $|A_i(t)| \le D$. The formal functional consumes the tuple of numerical states at those indices. The maximum product norm is

$$\|x - y\|_{\max} = \max_{j \in A_i(t)} \|x_j - y_j\|_2.$$

When $A_i(t)$ is empty, the product domain is a singleton and its input distance is zero. The target coordinate is still read and written, so an empty additional-input set does not make the update cost zero.

Validated temporal conditions may parameterize $f_{i,t}$. The notation t includes the chosen realization for purposes of the fixed-regime analysis. Comparing two trajectories under the estimates below requires holding those parameters, the selected indices, and the schedule fixed. If their acquisition depends on the numerical state, that dependency requires its own analysis before the estimates apply to the composite process.

The unit-ball state vector is precisely scoped to the numerical coordinates specified by this model. A full event history or arbitrary content is not an additional coordinate merely because it helps establish context. Similarly, an observation of a layer is not automatically the layer state itself.

The work theorem assumes that the active inputs can be presented in bounded form. It does not prescribe a fixed-capacity graph frontier, a graph search, or a history replay. Those would be implementation choices, not consequences of temporal locality. Their costs and semantics must be documented if adopted by a particular realization.

## II.4 Lipschitz coordinate maps

Let $f_{i,t}$ map the admissible active-input product into $\mathbb{R}^d$. Assume that, for all admissible x and y,

$$\|f_{i,t}(x) - f_{i,t}(y)\|_2 \le L_{i,t} \|x - y\|max,$$

with $\sup_{i,t} L_{i,t} \le L < \infty$. Fix $\eta > 0$ and set

$$T_{i,t}(\Psi) = \Pi(s_i + \eta\, f_{i,t}(\Psi|A_i(t))).$$

The generator is defined coordinatewise:

$$(G_{i,t}\Psi)_i = T_{i,t}(\Psi);\ (G_{i,t}\Psi)_k = s_k \text{ for } k \ne i.$$

This separates the prior target state from the local functional's selected inputs. The target may also be among those inputs if a particular functional requires it. The dependency union $A_i(t) \cup \{i\}$ still has cardinality at most D + 1.

The map is nonlinear in general. Accordingly, the relevant estimate is a Lipschitz inequality between two configurations. A multiplicative linear-operator estimate based on $\|\Psi\|_\infty$ is inappropriate without additional structure: a nonzero forcing term can produce a nonzero update even at $\Psi = 0$.

No global magnitude bound on $f_{i,t}(0)$ is needed for the state-invariance result because projection bounds the output. Regularity and numerical evaluation are separate matters. A production realization must still handle exceptional values and finite representation safely. Its numerical domain cannot be inferred solely from the exact-real formula.

## II.5 State invariance: local estimate

**Proposition 1.** Every generator defined in II.4 maps S into S.

**Proof.** Fix $\Psi \in S$ and target i. At the updated coordinate, the definition of $\Pi$ gives

$$\|(G_{i,t}\Psi)_i\|_2 \le 1.$$

For $k \ne i$, the generator leaves $s_k$ unchanged. Since $\Psi$ is admissible, each such coordinate already has norm at most one. Taking the supremum over all indices therefore gives

$$\|G_{i,t}\Psi\|_\infty \le 1.$$

Thus $G_{i,t}\Psi \in S$. This proves the proposition.

**Explanation.** The bound is independent of the total number of coordinates because no sum over them appears. It also does not depend on the number of prior transitions. Provided each successive map has the stated form, admissibility is maintained at every finite stage.

In a finite-dimensional implementation, invalid intermediate arithmetic can prevent the formula from being evaluated. Rejection and unchanged-state behavior need explicit implementation rules. The engineering tests of rejection and atomicity reported later support particular production paths; they are distinct from this exact-real projection proof.

## II.6 Sensitivity of one transition

For $\Psi, \Phi \in S$, write $\Phi = (\varphi_k)_k \in V$. Unchanged coordinates satisfy

$$\|(G_{i,t}\Psi)_k - (G_{i,t}\Phi)_k\|_2 \le \|\Psi - \Phi\|_\infty.$$

At i, nonexpansiveness of $\Pi$ and the triangle inequality imply

$$\|T_{i,t}(\Psi) - T_{i,t}(\Phi)\|_2$$

$$\le \|s_i - \varphi_i\|_2 + \eta\, \|f_{i,t}(\Psi|A_i(t)) - f_{i,t}(\Phi|A_i(t))\|_2$$

$$\le (1 + \eta L_{i,t})\, \|\Psi - \Phi\|_\infty.$$

Taking the maximum over the updated and unchanged coordinates gives

$$\mathrm{Lip}(G_{i,t}) \le 1 + \eta L_{i,t} \le 1 + \eta L.$$

This bounds the amplification of a perturbation for one map with a fixed context. The estimate is generally greater than one and therefore does not establish contraction.

If a particular $f_{i,t}$ is a sum of at most D interaction terms, each Lipschitz with constant at most $L_0$ in the same maximum input norm, the triangle inequality gives $L_{i,t} \le DL_0$. That is an optional sufficient structure for the bound, not a prescribed engine formula.

The hypothesis concerns the entire local functional, including any state-dependent inputs it actually evaluates. A constant proved while holding an external signal fixed cannot be reused for a map that recomputes that signal from the evolving state without also analyzing that dependence.

## II.7 Finite modification and bounded increments

Define the increment at the target by

$$\Delta_{i,t}(\Psi) = T_{i,t}(\Psi) - s_i.$$

Since the old and new target states are both in the Euclidean unit ball,

$$\|\Delta_{i,t}(\Psi)\|_2 \le \|T_{i,t}(\Psi)\|_2 + \|s_i\|_2 \le 2.$$

Every other coordinate of $G_{i,t}\Psi - \Psi$ is zero. Consequently,

$$\|G_{i,t}\Psi - \Psi\|_\infty \le 2.$$

This restores the original bounded-increment observation without calling the nonlinear generator a rank-one linear operator. It concerns the size and modification support of an increment, not the computational difficulty of finding it.

The distinction between dependency support and modification support is essential. A transition can inspect several states and still modify only one. Over multiple transitions, information can propagate through overlapping dependency sets. The support of an entire composition need not remain a single coordinate or a fixed subset as the number of transitions grows.

Nor does a single-coordinate write automatically ensure concurrency safety. Two computations can read an earlier state and attempt incompatible commits, or one can depend on a coordinate changed by another. A physical realization requires an ordering, isolation, or validation mechanism appropriate to its intended semantics.

The increment bound is independent of M, but it does not bound the total variation over an unbounded number of updates by a finite constant. Summing the bound over n updates gives at most 2n, which grows with n. Bounded state and bounded accumulated path length are different claims.

## II.8 Ordered composition and finite evolution

For a fixed schedule and active-set realization, define $g(0) = \mathrm{Id}$ and

$$g(t) = G_{\pi(t-1),t-1} \circ \cdots \circ G_{\pi(0),0}.$$

The composition of Lipschitz maps is Lipschitz. Applying II.6 successively gives

$$\mathrm{Lip}(g(t)) \le \prod_{k=0}^{t-1} (1 + \eta L_{\pi(k),k}) \le (1 + \eta L)^t.$$

Therefore, for two initial configurations using the same realization,

$$\|g(t)(\Psi_0) - g(t)(\Phi_0)\|_\infty \leq (1 + \eta L)^t \|\Psi_0 - \Phi_0\|_\infty.$$

Admissibility additionally bounds the left-hand side by two. The resulting finite-horizon estimate can be written as the smaller of these two bounds.

The conclusion is continuity of every finite composition with an explicit bound. It is not a uniform-in-time stability or convergence theorem. If $L > 0$, the displayed upper bound grows with t. Particular maps may have better constants, but those must be established for those maps.

The realization is held fixed in this comparison. If a state-dependent controller changes the selected contexts between the two trajectories, this composition estimate is no longer an estimate for their full closed-loop difference. Section III.4 treats that boundary explicitly.

This formulation retains the original temporal interpretation: local transitions accumulate into a global trajectory through ordered composition. It replaces the old linear-operator product language with a statement appropriate to nonlinear projected maps. General dynamical-systems context is supplied by [3, 4].

## II.9 Bounded algorithmic work

Regularity alone does not imply cheap evaluation. Assume explicitly that each $f_{i,t}$ has an evaluation procedure requiring at most

$$c_f^0 + c_f^1 |A_i(t)| d$$

primitive operations, with constants uniform in i, t, and M. Assume also that the active set is already available in bounded form and that coordinate access is represented by the chosen primitive-operation model.

Reading the target and active inputs requires at most $D + 1$ vectors. Fixed-dimensional addition, a Euclidean norm, unit-ball projection, and one coordinate write require a number of arithmetic operations proportional to d. Combining these costs yields constants $c_0$ and $c_1$ such that

$$C_{alg}(G_{i,t}) \leq c_0 + c_1(D + 1)d.$$

For fixed D and d, this is $O(1)$ with respect to M. The target term ensures that the estimate remains meaningful when $D = 0$. Constants do not depend on global cardinality under the stated uniform assumptions.

The estimate is for one application, not the acquisition of an arbitrary context. An unbounded adjacency scan, a global geometry refresh, a growing receipt lookup procedure, or replay of all prior events must not be silently counted as a constant primitive if that cost model is not justified.

Nor does the estimate say that a sweep over M targets is constant work. M applications can require work proportional to M. Persistence and index construction have separate input sizes. The useful consequence is that expanding retained state need not increase the primitive work of a specified local application whose assumptions remain fixed.

## II.10 From work to physical latency

A primitive-operation bound does not control hardware by itself. To obtain a physical-time bound, suppose the realization admits a bound $\tau$ on the time per counted primitive, a bound $C_{access}$ on additional state-access costs, and a bound $C_{exec}$ on execution delays not already included. Then

$$T_{update} \le \tau[c_0 + c_1(D + 1)d] + C_{access} + C_{exec}.$$

The decomposition must avoid double-counting: if an access cost is already included in $\tau$, it is not charged again. Each bound refers to a specified hardware, residency, concurrency, and workload regime.

The additional terms may include waiting for a lock, scheduling delay, page faults, memory compression, remote transfer, or other costs outside the chosen arithmetic model. They are not bounded independently of M merely because the numerical map has finite support.

Possible realizations include resident working sets, bounded partitions, controlled remote access, and placement suited to the access pattern. These are engineering options, not mandatory implementations of the temporal thesis. Spatial placement can improve access, but logical temporal locality and physical placement are not identical concepts.

Thus BLGC separates global cardinality, incremental algorithmic work, and physical execution conditions. The first can grow without increasing the second under the formal assumptions. Whether the third remains stable is an experimental and systems question. The measurements in Section IV address selected regimes; they do not establish a universal latency ceiling.

# III. Finite-horizon guarantees

## III.1 Admissible-state preservation

**Theorem 1.** Let $\Psi_0 \in S$. Suppose every transition has the form defined in II.4. For every finite $t \geq 0$, $\Psi_t = g(t)(\Psi_0)$ belongs to S, and $\|\Psi_t\|_\infty \leq 1$.

**Proof.** At $t = 0$ the claim is the initial-state assumption. If $\Psi_t \in S$, Proposition 1 gives $G_{\pi(t),t}(\Psi_t) \in S$. Induction proves the claim for every finite t.

**Interpretation.** Increasing the number of stored coordinates does not change the numerical bound on each coordinate. Nor does the proof require square summability across all indices. The same argument applies to every finite restriction with compatible transition definitions and to the countable bounded-sequence model.

The theorem is conditional on the transition form. An additional unprojected mutation does not inherit the conclusion simply because it occurs in the same subsystem. Migration, restoration, event impulses, and exceptional-value handling require their own domain-preservation arguments if they are to be included in a complete implementation theorem.

## III.2 Finite-horizon perturbation control

**Theorem 2.** Under the uniform Lipschitz assumption, two admissible initial configurations evolved through the same sequence of maps satisfy

$$\|\Psi_t - \Phi_t\|_\infty \leq \min\{2, (1 + \eta L)^t \|\Psi_0 - \Phi_0\|_\infty\}.$$

**Proof.** Apply the single-map estimate successively to obtain the composition bound in II.8. Independently, both configurations belong to S by Theorem 1, so their distance is at most two. Taking the smaller bound proves the result.

The first bound is informative about sensitivity over a specified finite horizon; the second records confinement to the admissible set. Neither says that the separation decreases. A bounded trajectory can retain or amplify small differences before reaching the domain-size limit.

For nonuniform local constants, the product of the actual factors can replace $(1 + \eta L)^t$. This may give a sharper statement, provided the same fixed realization is used for both trajectories and the constants are valid on the relevant domains.

The theorem does not compare runs with different input events, schedules, policy histories, or selected context. Such differences alter the maps themselves. A comparison involving them needs explicit bounds on those changes; they cannot be folded into the initial-state distance without justification.

Theorem 2 establishes finite-horizon perturbation control for the stated realization. The growth factor and the common-map hypothesis determine its scope.

## III.3 Dimension and incremental-work decoupling

**Theorem 3.** Consider a family of finite instances with M coordinates, fixed d and D, and uniform evaluation constants as in II.9. One generator application has a primitive-work bound independent of M:

$$C_{alg}(G_{i,t}) \le c_0 + c_1(D + 1)d.$$

**Proof.** At most D additional input vectors and the target are required. The assumed functional evaluation has a uniform cost. Addition, Euclidean projection, and one coordinate write contribute O(d) primitive operations. Combining these bounds gives the result without a term in M.

The total finite coordinate dimension dM can increase while this bound remains unchanged. This is dimension-work decoupling for one defined transition. The substantive implementation obligation is the uniform acquisition and evaluation procedure assumed in II.9; a small result vector alone does not discharge it.

If the physical assumptions of II.10 also hold uniformly across the instance family, the corresponding physical-time bound is likewise independent of M within that regime. Without them, the theorem remains an algorithmic-work result.

## III.4 Switching regimes and discrete selection

A regime fixes the numerical domain, the selected dependency sets, the relevant external context, and the maps being composed. State-dependent selection can move the system between regimes. A changed topology, policy, dimension, or signal realization may require a different regularity or work analysis.

Theorem 2 applies within a common fixed realization. A deterministic controller does not, by itself, extend it across switching boundaries. For instance, an input selected by a ranking rule can change discontinuously when the ranking changes. Deterministic tie-breaking removes ambiguity at equality, but does not establish continuity across that boundary.

One legitimate route is to analyze each regime and then specify an admissible switching rule with its own estimates. Another is to prove continuity of the complete controller-and-state map. Neither route is assumed completed here for the full production engine.

This distinction fits the temporal thesis. Successive state transitions need not erase their past simply because the applicable policy changes; the applicable history and boundary must be known. At the same time, a numerical constant established for one policy cannot silently stand for another.

## III.5 Discrete realization and finite arithmetic

The formal maps act on exact real vectors at discrete transition indices. An implementation uses finite representations and numerical procedures. Domain checks, overflow behavior, rounding, and projection evaluation must therefore be considered separately from the exact-real proof.

A standard conditional error statement clarifies what would be needed. Suppose an approximate step introduces an error bounded by $\varepsilon_k$ in the same norm, and the exact map at step k has Lipschitz constant $a_k$. Then errors satisfy

$$e_{k+1} \leq a_k e_k + \varepsilon_k.$$

Iterating gives

$$e_t \leq (\prod_{k=0}^{t-1} a_k)e_0 + \Sigma_{j=0}^{t-1} \varepsilon_j(\prod_{k=j+1}^{t-1} a_k).$$

Empty products equal one. The inequality follows by substituting the previous-step bound repeatedly. It becomes a certified implementation bound only when the local errors and constants have been justified for the actual arithmetic, inputs, and horizon.

This paper does not claim such a certified full-engine floating-point bound. Nor does a passing partition test certify every possible time interval, input magnitude, or numerical library. Focused tests can detect defects and establish concrete examples while leaving the universal proof obligation explicit.

The distinction also prevents a discretely indexed map from being described as a proven integration scheme for an unspecified differential equation. The public BLGC construction is the projected map stated above. A separate continuous-time model and its numerical realization would need their own hypotheses and error analysis.

# IV. Implementation and evidence

## IV.1 From temporal organization to engine evidence

OPAL is the persistent-state engine; KINDRED provides an application and integration context. Each transition uses an identifiable prior state, defined input and accounting context, an authorized mutation, and a result available to later operations. Chronology, ownership, replay and recovery preserve this continuity through execution and restart.

Tests verify typed chronology, separate passive and reinforcing access, replay/conflict handling, atomic state-and-receipt commits on exercised paths, and continuity across publication and recovery. IV.2 summarizes these reliability results.

The operational bridge is established through persisted state and tested ordering and recovery contracts. The correspondence register below connects that evidence to the formal requirements for a projected numerical transition with bounded dependencies and evaluation.

## IV.1 continued: correspondence register

| Formal requirement | Relevant recorded engine evidence | Remaining correspondence obligation |
| --- | --- | --- |
| Fixed-dimensional $s_i$ in a specified admissible domain | Two-coordinate resident substate, explicit context and admission domain traced (IV.11) | Accumulating production coordinates and projected formal coordinates have distinct domains; see IV.18 |
| A defined transition from carried state and context | Production owner transition traced; canonical outcomes match before and after index repair | Full layered-state correspondence remains distinct from this scoped transition |
| At most D additional inputs, with bounded acquisition | One-target evaluation; incremental receipt indexing removes repeated whole-index reconstruction | Hash access and capacity growth retain expected/amortized cost qualifications |
| Uniform functional evaluation and coordinate update | Initial nine-run study plus 18 matched processes directly measure the owner boundary | Fixed policy/context, record shape and cache regime define the measurements |
| A common schedule/context realization for sensitivity analysis | Fixed-input chronology/replay tests and generation-based rejection on covered paths | Prove the required regularity for the mapped state and analyze any state-dependent selection |
| State continuity through persistence | Controlled publication failure, rollback/retry, and restart/recovery tests | Establish correspondence for every state component and transition included in a full implementation theorem |

The register identifies state, inputs, mutations and costs for the implementation claim. The studies complete this trace for one resident transition and validate its acquisition repair; IV.18 consolidates the remaining formal scope.

## IV.1 continued: carried state and retained history

A target supplies resident state and accounting context; an explicit operation commits a change; the next operation continues from that result. Recovery restores the state and identity information needed for continuation and replay.

Retained history has a different role. It may support provenance, audit, recovery, or authorized reconstruction. Its existence does not require that each resident numerical step scan it. Conversely, any path that does scan history must charge that work to the operation that performs it. The same accounting applies to index rebuilding, global Flow/geometry processing, and checkpoint publication.

Earlier changes remain represented in carried state while historical evidence follows separate lifecycle contracts. The tests establish continuity and recovery; application-specific evaluation determines which useful information the state retains.

The study measures both semantic preservation through use, replay and restart, and the work of each named operation. IV.11-IV.17 report the production mapping and controlled experiments; V.2 summarizes the results.

## IV.2 Operational reliability: preservation, replay and recovery

**The engine passed the tested reliability contracts.** Reliability here means preserving accepted state and event identity through normal execution, invalid input, retries, concurrent calls, storage growth and restart. The evidence comes from explicit state checks and canonical comparisons, alongside successful workloads.

**Temporal accounting and access.** Typed chronology separates revision from time. Eligible influence evolution uses a consistent accounting origin across tested maintenance, event, and restart boundaries. Passive reads preserve resident state. Explicit reinforcing access is distinguished from observation.

**Events and ownership.** Identity-bearing events support replay and conflict detection. Relevant state changes and receipts commit atomically on the exercised paths. Project isolation prevents another runtime's reinforcement history from independently affecting the tested scoring path. Identity-less compatibility calls do not gain an exactly-once guarantee by assumption.

**Preservation and recovery.** Missing legacy evidence is retained as unavailable rather than fabricated. Protected authoritative state is preserved under the tested policy. Publication failure exercises exact rollback and retry. Recoverable archive and directed-link lifecycle operations have separate authorization and recovery boundaries; permanent deletion remains disabled.

**Persistence and concurrent access.** Controlled restart tests verify the recorded temporal and replay behavior. Concurrent query fixtures return coherent committed results under the tested operations. Successful in-memory acceptance and successful durable publication are distinguished.

**Matched reliability evidence.** The receipt-index study produced byte-identical original and repaired checkpoints and matching continuation after independent-process recovery. Eight concurrent callers submitting one identity produced one commit and seven replays. Rejection, legacy receipt compatibility, allocation-failure fallback and capacity-growth checks passed (IV.15). All 18 matched scaling processes passed their state, chronology, integrity, replay and conflict gates (IV.16); nine forced-growth processes also passed (IV.17). These are demonstrated operational outcomes, not conclusions inferred from fast timings.

## IV.3 Final scale methodology

The final scale evidence comprises three fresh sequential Release processes at one million nodes and three at five million nodes. Every report records a fresh process. The platform was an Apple M2 Pro with 16 GB of memory, arm64, macOS 26.6.2, using the recorded Release toolchain. All six processes completed successfully and passed the declared integrity and traversal checks.

The workload constructs a synthetic graph, performs persistence save/load, refreshes global Flow/geometry telemetry, exercises reinforcing indexed access, and checks traversal behavior. It is not a comprehensive temporal-event sweep or an experiment measuring semantic retrieval quality.

For indexed access, the method selects 100 deterministic IDs. Each ID receives one warmup and 100 timed accesses. Its timed durations are averaged, producing 100 per-ID means. The reported percentiles are taken across those means. They are not percentiles across all individual calls.

The percentile recipe selects the sorted index $\mathrm{floor}[p(n - 1)]$. With $n = 100$, P50 uses index 49. The separately labeled median uses index 50. Small differences between the median and P50 fields are therefore expected under this report convention; they are not silently reconciled by renaming one as the other.

The access operation is in-process and reinforcing. It is not a network request, a semantic nearest-neighbor search, or a complete graph traversal. Persistence is captured before that lookup phase. The synthetic fingerprints cover the declared graph recipe and invariants, not every possible field of the entire resident object graph.

The two scales use different degree/topology regimes. The results demonstrate successful execution at both sizes, rather than an isolated causal law in M.

## IV.4 Final indexed-access results

**Table 1.** Microseconds across per-ID mean reinforcing indexed-access durations. Each row is a separate fresh Release process; percentile and median conventions are defined in IV.3.

| Run | Median | P50 | P95 | P99 |
|---|---|---|---|---|
| 1M / 1 | 0.12541 | 0.12541 | 0.16213 | 0.23207 |
| 1M / 2 | 0.12667 | 0.12666 | 0.16585 | 0.17542 |
| 1M / 3 | 0.12750 | 0.12709 | 0.13124 | 0.13208 |
| 5M / 1 | 0.12915 | 0.12874 | 0.13541 | 0.15169 |
| 5M / 2 | 0.12626 | 0.12626 | 0.13207 | 0.15290 |
| 5M / 3 | 0.12707 | 0.12668 | 0.13334 | 0.13626 |

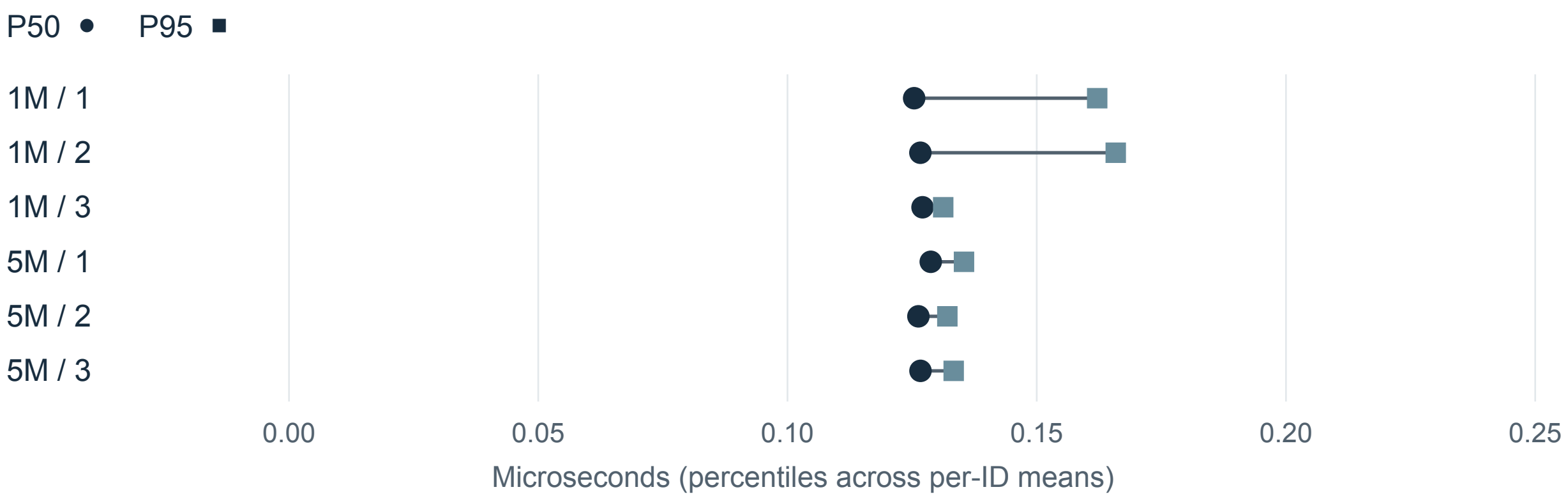

**Figure 1.** P50 and P95 observations from Table 1, on a common time axis. Each mark represents a reported percentile across 100 per-ID averages. The figure displays the six observed runs; it is not a fitted scaling curve or a confidence interval.

Across these runs, P50 lies between 0.12541 and 0.12874 microseconds. The result is repeatability of this warm, in-process indexed-access measurement under the declared workload. It does not establish individual-call tail latency, semantic-search quality, or constant end-to-end cost for all engine operations.

## IV.5 Memory, persistence, and growth

**Table 2.** Sampled resident-set sizes in decimal GB, where 1 GB = $10^9$ bytes. Values are maxima among recorded samples, not continuous peaks or object allocation totals.

| Run | Persistence-stage maximum | Whole-workload maximum |
|---|---|---|
| 1M / 1 | 4.065346 | 4.065346 |
| 1M / 2 | 3.025043 | 3.025043 |
| 1M / 3 | 3.843981 | 3.843981 |
| 5M / 1 | 5.290197 | 6.215483 |
| 5M / 2 | 7.627162 | 8.420590 |
| 5M / 3 | 7.627162 | 7.627162 |

The whole-workload range was 3.025043-4.065346 GB at one million nodes and 6.215483-8.420590 GB at five million. In binary units these ranges are approximately 2.817291-3.786148 GiB and 5.788620-7.842285 GiB. A general claim that the complete five-million-node workload stays below 5 GB is not supported.

Persistence size, resident memory, allocator reservation, and live allocated bytes are different measurements. A compact serialized file does not establish an equally small runtime footprint. Conversely, a higher resident-set sample does not identify which objects are responsible.

The synthetic records also do not represent every possible user payload. Variable-length content, relationships, audit evidence, and retained archives can change storage requirements. Growth of those domains is not constrained by the unit-ball numerical theorem.

## IV.6 Historical access results and label reconciliation

Historical access results use the same per-ID-average convention, but their source/runtime provenance and workload differ from the final six-run series and original screenshot.

**Table 3.** Historical indexed-access durations in microseconds.

| Historical scale | Median | P50 | P95 | P99 |
|---|---|---|---|---|
| 1M | 1.11207 | 1.11205 | 1.23874 | 1.29335 |
| 5M | 1.20877 | 1.20754 | 1.40751 | 1.50421 |

The original approximate “1.4 microseconds median” language is corrected: the historical five-million value 1.40751 microseconds is P95 across per-ID means. The distinct median and P50 fields are retained rather than merged.

These historical graphs had 2,999,889 and 24,988,889 logical edges, with maximum degrees six and ten. The original Figure 1 screenshot instead displays 14,999,889 edges for its five-million example. That difference prevents treating the screenshot and this table as one reformatted dataset.

The final timings are smaller than these historical timings, but this paper does not assign a speedup to the repairs from that comparison. Source, runtime state, and other conditions would need a matched experiment for causal attribution. A separate matched SDK comparison is reported in IV.10 with its own limitations.

These values retain their indexed-access and per-ID-average definitions; they are not traversal, network or individual-call latency measurements.

## IV.7 Historical traversal and persistence scope

Traversal evaluates a path rather than simply returning an indexed record. The archived report distinguishes a first-pass application-cache-miss case from a separate resident-cold case. Those labels are not claims that hardware caches or the operating system were comprehensively flushed.

**Table 4.** Historical three-hop traversal durations in microseconds, as labeled in the archived evidence.

| Case | Scale | P50 | P95 | P99 |
|---|---|---|---|---|
| First-pass application-cache miss | 1M | 33.084 | 136.458 | 235.833 |
| First-pass application-cache miss | 5M | 44.250 | 77.375 | 85.083 |
| Resident-cold | 1M | 4.542 | 9.125 | 12.000 |
| Resident-cold | 5M | 7.833 | 11.375 | 13.916 |

The historical serialized sizes were 610,857,411 bytes and 3,166,222,153 bytes. Reported save/load times were approximately 2.19856/2.28365 seconds at one million nodes and 18.0383/20.8711 seconds at five million. These are global persistence operations, not bounded local transitions.

Pre-save and post-load fingerprint agreement supports the declared structural checks. A single serialized-file hash does not establish equality across independently executed replays. Likewise, a graph fingerprint's coverage must be known before treating it as a fingerprint of every numerical field and history.

The original screenshot's CPU plots and timing labels are retained in Appendix A for historical continuity. They are not used to infer tight jitter, linear hardware scaling, or current temporal-transition performance. This preserves the record without combining incompatible workloads.

## IV.8 Complete queries during maintenance

A separate Release fixture used 513 nodes and 512 candidates. Four fresh processes each ran 2,000 complete queries in four cases: idle, concurrent temporal events, concurrent Flow refresh, and concurrent archive restoration. The maintenance workloads comprised 128 events, four Flow calls, and one restoration operation, respectively.

**Table 5.** Ranges across four runs, in microseconds. These are complete-query measurements, distinct from Table 1's indexed-access averages.

| Case | P50 range | P95 range | P99 range |
|---|---|---|---|
| Idle | 43.459-50.875 | 43.750-89.583 | 48.583-106.541 |
| Temporal events | 51.959-52.583 | 53.083-54.750 | 54.583-58.458 |
| Flow refresh | 51.958-52.458 | 53.292-53.833 | 54.833-57.375 |
| Archive restoration | 51.917-52.208 | 52.834-53.708 | 54.625-57.583 |

All 32,000 measured queries succeeded and passed relationship-membership and eligibility checks. Maintenance succeeded. In each concurrent case and run, 2,000 queries overlapped the complete maintenance-operation interval.

This is operation-overlap evidence. Overlap with the restoration publication lock was not observed, and the interval measurement does not prove continuous maintenance-worker progress during every query. Validation outside timed query regions can still affect scheduling and overall throughput.

## IV.9 Allocation observations and attribution

Memory analysis used a bounded fixture of 2,048 AMVs with 64-component embeddings. Three fresh Release processes produced repeatable phase-boundary default-allocator-zone observations. The canonical checkpoint measured 1,360,696 bytes.

Reported default-zone bytes in use were 106,320 at baseline; 2,514,784 after retaining the fixture; 2,515,856 after encoding; 5,039,360 after decode, runtime installation, and index construction; and 5,041,408 after Flow completion. These are observations of that allocator zone, not object-level attribution or every allocator zone in the process.

The 2,048-byte difference after Flow completion is a retained boundary delta. It does not identify the transient high-water mark. Allocator reservation and sampled RSS varied despite stable bytes-in-use observations, illustrating why these measurements must remain separate.

An authorized allocation recording subsequently identified Flow snapshot and PCA/geometry allocation stacks. The filtered tree reported approximately 1.42 MB across 2,114 allocations. These cumulative allocation observations are not peak live memory or RSS. A “created and destroyed” category establishes release during the recording, not each object's exact destruction time.

No production optimization was introduced merely to obtain a different number. The evidence supports a cost account and identifies allocation activity in the bounded fixture. It does not attribute the historical five-million-node memory footprint, prove that no optimization opportunity exists, or establish a constant bound on whole-graph Flow work.

## IV.10 Matched SDK compatibility workloads

A supplementary comparison used the recovered, hash-matched archived SDK runtime and the repaired runtime on the same machine with Python 3.14.6. The local workloads were DCPerf-oriented; this is not a claim of upstream DCPerf certification or a networked data-center deployment.

Each workload used 100 warmups and 1,000 measured operations in a fresh child process, with unchanged workload parameters. Lookup used 128 states, traversal 1,200 states, recall 96 states, and ingestion eight states per call. The traversal fixture had 3,489 edges and verified application-cache misses and connected paths. Recall queried node IDs, not semantic relevance.

**Table 6.** Original matched comparison in operations per second: one archived-runtime run and the first repaired-runtime run.

| Operation | Archived rerun | Repaired runtime | Difference |
|---|---|---|---|
| Lookup | 364,370.53 | 373,407.16 | +2.48% |
| Ingestion calls | 24,865.67 | 26,751.14 | +7.58% |
| Connected traversal | 65,875.25 | 64,880.82 | -1.51% |
| Node-ID recall | 60,047.74 | 60,576.23 | +0.88% |

All four correctness gates passed. The repaired engine retained comparable performance across the exercised workloads while preserving lookup, state-and-relationship ingestion, traversal, and node-ID recall correctness.

These timings include Python and harness overhead and use the existing whole-loop timing method. They must not be compared directly with Table 1 as though the operations were identical. The original matched pair is descriptive, not a confidence interval or statistically reliable improvement/regression. Compilation occurred between that pair of runs;

environmental variation and archived compiler provenance limit causal attribution. The second current-engine run is reported separately in Table 7.

## IV.10 continued: repeat run of the repaired SDK

A second run on the Apple M2 Pro laptop with 16 GB RAM used the same repaired Release extension, verified by its SHA-256 identity. CPython 3.14.6, macOS 26.6.2 arm64, a single Python thread, 100 warmups and 1,000 measured operations per workload were retained. Each workload ran in a fresh child process. No rebuild, profiler, source change, or workload change accompanied this run.

**Table 7.** Second current-engine SDK run. Throughput is operations/s (ingestion calls/s); latency is in microseconds. All four workloads passed.

| Workload | Throughput | Mean | P50 | P95 | P99 |
|---|---|---|---|---|---|
| Indexed lookup | 377,631.53 | 2.579 | 2.542 | 2.875 | 3.333 |
| Text ingestion | 25,382.46 | 39.286 | 34.542 | 53.542 | 104.084 |
| Connected traversal | 68,300.96 | 14.584 | 14.042 | 17.667 | 24.000 |
| Node-ID recall | 62,016.47 | 16.040 | 15.417 | 18.750 | 24.209 |

Ingestion inserted eight states per measured document, yielding 203,059.64 states/s and 8,000 measured states, in addition to warmup states. Lookup returned every requested ID. Every traversal was an application-cache miss and returned a connected path of at least two nodes. Every recall returned its expected node ID; this remains a direct-ID workload, not a semantic-quality test.

Throughput uses the whole-loop timer; mean and percentile latencies are diagnostic per-call samples, not reciprocals of whole-loop throughput. These SDK samples include the public call boundary and remain separate from Table 1's native percentiles across per-ID means.

Relative to the preceding current-engine run, throughput changed by +1.13% for lookup, -5.12% for ingestion, +5.27% for traversal, and +2.38% for recall. The second successful run adds repeat-execution evidence with the same binary. Two current-engine runs do not establish a statistically reliable speedup, regression, or distribution of ordinary variation. This remains local DCPerf-style testing, not independent Meta/DCPerf validation.

## IV.11 Production state, domain and transition mapping

The study traces an existing resident event from its carried numerical state through explicit event/time inputs, stored context, applicable policy, validation, numerical evaluation, installation and mutation publication. The numerical substate has two coordinates. Chronology, identity, content, embedding, relationships and receipts retain their own typed representations and roles in the enclosing operation.

The evaluator reads one target substate and fixed context, with no traversal of other nodes. Under the formal active-set convention, the target may belong to $A_i(t)$; the distinct dependency union still contains one node. Private source analysis establishes nonexpansive exact-real sensitivity for a common fixed accepted context. This identifies a real local mechanism and its dependency boundary.

The production coordinates accumulate accepted input. Their applicable domain is finite representable numerical values subject to the existing admission checks. A fixed finite-horizon input budget supports a corresponding bounded-state analysis; unlimited accepted accumulation does not supply a history-independent unit-ball bound. This representation preserves the engine's intended meaning. No projection, normalization or change to the numerical law is introduced.

The work trace separates evaluation from acquisition and publication. The original receipt path rebuilt a derived ID index after each successful receipt append. The repair in IV.15 maintains that index incrementally under the same owner lock. Initial index construction, policy-history evaluation and pending cache work remain identifiable costs. The formal domain relationship is recorded in IV.18; the following experiments measure the actual production operations.

## IV.12 Controlled cardinality study: method and correctness

The initial experiment used the unchanged frozen production core, built in native arm64 Release with Apple clang 21.0.0, C++17 and -O3 on an Apple M2 Pro with 16 GB RAM, macOS 26.6.2. Nine sequential fresh processes covered three repetitions at each of 10,000, 100,000 and 1,000,000 data nodes, with scale order counterbalanced across repetitions. One caller thread performed the operations; the runtime worker-count setting was one. No profiler was active.

The graph family was a directed ring with two successor links per data node. Record shape, 13-byte IDs, eight-component embeddings, the two-coordinate numerical substate, fixed policy, stored context and event schedule were held constant. Each process used 1,024 evenly distributed targets in the same deterministic permutation. The whole graph necessarily changes with node count; degree and local topology pattern remain fixed. No policy-history changes, Flow recomputation, populated traversal cache, or concurrent mutations occurred.

After untimed construction, index preparation and one warmup visit per target, each primary operation received 20 measured visits per target: 20,480 calls per process. Passive access included copying the fixed-shape record. The resident operation included target resolution, validation, temporal evaluation, commit and generation/cache publication. Explicit one-second intervals and alternating small signed contributions exercised real transitions. Every call had to succeed, and all 1,024 final target states and their chronology/content/topology invariants were checked outside timing against a separate numerical reference; an unvisited node was also checked.

A prior bounded validation passed accepted-state checks, stale-time/nonfinite/policy rejection atomicity, passive-read purity, receipt replay/conflict, and the domain counterexample. A genuinely new process recovered its checkpoint, replayed the saved identity and continued the resident transition. These are native owner tests. They add no Python-package or cross-platform certification.

Timing used steady_clock per call and a separate whole-loop timer. P50/P95/P99 use nearest rank over individual calls, unlike Table 1's per-ID means. The protocol, raw samples, source manifest, executable identity, process records and checks are retained. Setup and index preparation are outside the primary spans and explicitly recorded.

## IV.13 Controlled resident-update results

All nine initial-study processes passed. Each primary operation contributed 184,320 measured calls across the study. Table 8 reports individual-call microseconds. Each row is a fresh process; the resident operation is the actual owner update rather than an isolated arithmetic helper.

**Table 8.** Controlled fixed-regime measurements, in microseconds. Passive P50 includes record copy; the remaining columns describe the resident event update.

| Run | Passive P50 | Resident P50 | Resident mean | Resident P95 | Resident P99 |
|---|---|---|---|---|---|
| 10k / 1 | 0.083 | 0.083 | 0.086 | 0.125 | 0.125 |
| 10k / 2 | 0.083 | 0.083 | 0.087 | 0.125 | 0.125 |
| 10k / 3 | 0.083 | 0.083 | 0.093 | 0.167 | 0.209 |
| 100k / 1 | 0.083 | 0.083 | 0.092 | 0.125 | 0.125 |
| 100k / 2 | 0.083 | 0.083 | 0.087 | 0.125 | 0.125 |
| 100k / 3 | 0.083 | 0.084 | 0.111 | 0.209 | 0.458 |
| 1M / 1 | 0.125 | 0.084 | 0.108 | 0.125 | 0.209 |
| 1M / 2 | 0.125 | 0.084 | 0.104 | 0.125 | 0.167 |
| 1M / 3 | 0.084 | 0.084 | 0.101 | 0.125 | 0.167 |

Resident-update P50 stayed at 0.083-0.084 microseconds in this regime. These short durations are close to clock quantization: empty timer pairs had P50 zero and P95 42 nanoseconds in every process. No timer subtraction was applied. A one-nanosecond difference between reported medians is not evidence of a speed change. Means and upper quantiles vary, as shown; flat medians do not imply identical latency distributions.

Separately timed whole-loop resident throughput ranged from 7.735 to 9.505 million calls/s and includes measurement bookkeeping. It is not the reciprocal of the reported percentiles. Peak process RSS ranged from approximately 0.021 GB at 10,000 nodes to 1.458-1.461 GB at one million, under this particular small-record fixture. Those values do not replace the earlier workload's memory results.

## IV.14 Original receipt-bearing boundary

In the initial unmodified-core study, after the primary measurements each process submitted 16 distinct identity-bearing events. The first began with a prepared index; each subsequent call followed a receipt append that invalidated that index. This diagnostic preserves the production operation, including its acquisition, validation, evaluation, receipt insertion and commit. Node count grows from M to M+16 during this phase.

**Table 9.** Ranges across three fresh processes. The first event is a single clean-index observation per process. Subsequent P50 is computed over the remaining 15 calls per process; these small samples do not characterize production tails.

| Data nodes | First event, µs | Subsequent P50, ms |
|---|---|---|
| 10,000 | 16.875-21.292 | 0.427-0.499 |
| 100,000 | 14.084-17.750 | 7.792-9.377 |
| 1,000,000 | 17.166-21.000 | 170.944-175.223 |

The substantial growth is consistent with the traced index reconstruction over retained records. These are whole-operation measurements, not a profile attributing every elapsed nanosecond to a particular component. Successful replay/conflict checks and final-state assertions remained outside the timed spans.

Tables 8 and 9 isolate an acquisition problem around a local numerical transition. The repeated reconstruction is a derived-index maintenance cost. The next study repairs that cost and compares both versions under the same workload; the original observations remain here as the historical diagnosis.

## IV.15 Incremental receipt indexing and semantic preservation

A receipt-bearing event appends an immutable identity record. The original path invalidated the full derived ID index although append preserves existing record positions. The repair inserts the new identity and position under the same lock, preserving first-occurrence and long-term-before-short-term lookup precedence.

Only receipt-append index maintenance changes. Numerical evaluation, admission, canonical payload, replay/conflict decisions, chronology, usage, generation publication and durable encoding retain their behavior. Erasure, reordering, replacement and restoration keep their existing index paths.

If incremental insertion cannot allocate, the commit completes with a dirty cache and the next lookup uses the existing rebuild fallback. Successful insertion leaves a complete index without visiting prior records.

Both versions passed public-owner checks for accepted transitions, stale/nonfinite/policy rejection, replay and conflict, legacy receipt compatibility, usage accounting and short-term targets. Separate fixtures verified duplicate-ID precedence and concurrent equal identities: one commit and seven replays. Baseline and repaired canonical checkpoint bytes matched exactly, including state continued after independent-process recovery. Failed checkpoint publication followed by retry preserved the same bytes.

Fixed-version cache tests additionally exercised entry and bucket allocation failures, existing dirty state, empty and duplicate IDs, vector allocation failure, and 200 successful receipt appends across capacity growth with a complete clean index. These tests validate the repair's behavioral contract without altering the state law.

## IV.16 Matched controlled scaling after the repair

The comparison used the same production source snapshot except for the receipt-index repair, the same Release settings, and the unchanged primary benchmark fixture from IV.12. Eighteen fresh processes covered three repetitions at each scale for each version. Baseline and repaired runs were interleaved, alternating which version ran first; scale order was counterbalanced. Each version contributed 184,320 resident-update calls, 184,320 passive calls and 144 receipt events. Every run passed final-state, chronology, record-integrity, replay and conflict checks.

**Table 10.** Ranges of per-process medians across three repetitions, in microseconds. Receipt P50 uses calls 2-16 of each process; the first call is separated. Both receipt columns measure the complete public transaction.

| Data nodes | Original receipt P50 | Repaired receipt P50 | Repaired resident P50 |
|---|---|---|---|
| 10,000 | 422.167-454.708 | 5.166-6.459 | 0.083-0.083 |
| 100,000 | 8743.042-9897.792 | 4.875-5.459 | 0.083-0.084 |
| 1,000,000 | 177699.667-182297.375 | 4.458-4.708 | 0.084-0.084 |

Repaired receipt medians show no increasing trend with node count in this workload. At one million nodes, the original 177.700-182.297 milliseconds fell to 4.458-4.708 microseconds. Resident-update medians remained 0.083-0.084 microseconds; IV.13 states the timer-resolution qualification. Together with the correctness gates, this supports the removal of repeated reconstruction while preserving the tested transaction behavior.

First repaired receipt durations, in microseconds, were 10,000: 17.417-18.584; 100,000: 15.542-21.334; 1,000,000: 19.333-23.333. All individual samples, build/source identities, process order and executable hashes are retained. The 15-call receipt medians characterize this bounded comparison; capacity-boundary measurements follow separately.

## IV.17 Capacity growth and the operation-cost boundary

Vector and hash-table growth remain occasional costs after incremental indexing. Nine additional fresh repaired-version processes, three per cardinality, measured those boundaries separately.

Untimed fixture preparation placed the vector at exact capacity, then timed one real receipt event that grew it. A second setup placed the hash table at its growth threshold, then timed another real event. Each process verified actual capacity growth, a complete clean index, replay and final numerical state. These are deliberately forced boundary observations, separate from the ordinary workload in Table 10.

**Table 11.** Ranges across three processes, in milliseconds. One event of each boundary type per process; setup is excluded.

| Data nodes | Vector-growth event | Hash-growth event |
|---|---|---|
| 10,000 | 0.760-1.136 | 0.140-0.155 |
| 100,000 | 7.439-9.133 | 1.167-1.511 |
| 1,000,000 | 189.330-251.888 | 70.486-79.657 |

The repair eliminates a whole-index rebuild on every ordinary receipt append; it does not eliminate container growth. This distinguishes the observed steady-state improvement from a worst-case constant-latency guarantee. Other receipt-producing operations, policy-history scans and cache maintenance retain their existing work and require their own operation-level accounting.

## IV.18 Consolidated scope, limitations, and evidence availability

The reliability tests establish the operational outcomes in IV.2 and IV.15-IV.17. The mathematical bounds apply under their stated hypotheses. The following qualifications define the reach of both results.

**State and correspondence.** The production substate in IV.11 accumulates accepted reinforcement and is not the projected unit-ball coordinate of Sections II-III. An accepted owner event takes its Euclidean norm from 0.25 to approximately 3.010399. This is valid production behavior. A finite-horizon input budget can bound that substate, but complete original layered-state correspondence and unlimited-history invariance are not established. Content, provenance, topology and retained history have separate representations and costs.

**Regularity and realization.** The formal sensitivity result compares common maps, context and schedule. Full state-dependent selection, cross-regime regularity, accumulated floating-point error and cross-platform bitwise identity require separate analysis. The tests establish the named replay and recovery outcomes.

**Work and measurement.** The controlled studies fix local topology, record shape, policy, stored context and cache regime. Hash access retains expected/amortized qualifications; Table 11 exposes capacity growth. Initial construction, maintenance, global Flow processing and persistence are separately charged. Submicrosecond timing is coarse, and the small receipt samples do not establish production tails. Physical execution remains subject to II.10's conditions.

**Application value and evidence availability.** Semantic quality, energy savings, hardware extrapolations and comparative superiority were not measured. The references supply background rather than a completed priority review. Raw samples, canonical comparisons, source/build identities, process records and private correspondence analyses are retained. The paper publishes aggregate evidence and workload definitions; private implementation details are not disclosed, limiting independent reproducibility.

The operational result is concrete: matched canonical state and recovery behavior, successful million-node execution, and removal of repeated index reconstruction from the measured receipt path.

# V. Discussion and conclusion

## V.1 Architectural and hardware interpretation

A growing collection can retain history while each specified transition uses carried state and selected context. The mathematics states sufficient conditions; the production study supplies continuity, persistence and scaling evidence. IV.18 defines their scope.

Local updates, context refreshes, queries and checkpoints have different input sizes. Each must account for its acquisition and maintenance costs at the measured operation boundary.

The hardware question follows from that account. Once an operation's dependencies and work are known, cache residency, memory placement, bandwidth, synchronization, and execution delay can be evaluated as actual bottlenecks. Peak bandwidth alone does not predict engine throughput. The original 100 GB/s-to-3,400 GB/s comparison remains a historical extrapolation, not evidence of a 34-fold engine speedup.

On a consumer M2 Pro laptop, the engine completed the declared million-node workloads. The controlled comparison then removed repeated per-event index reconstruction while preserving canonical state and recovery behavior.

## V.2 What the completed study establishes

The study traces a resident transition, measures its complete owner boundary across controlled cardinalities, and repairs receipt-index acquisition without changing the numerical state law or receipt semantics.

**Reliability and performance are supported by separate checks.** Canonical checkpoints and post-restart continuation match byte for byte; replay, conflict and concurrency checks pass. All 18 matched scaling processes pass. Resident-update medians remain 0.083-0.084 microseconds, and repaired receipt medians span 4.458-6.459 microseconds with no increasing trend across the tested cardinalities.

The growth study locates the remaining capacity costs. Ordinary events no longer rebuild the full index; occasional storage growth and global maintenance retain their own work. The practical result combines state continuity with incremental acquisition.

IV.18 states the domain and generalization boundaries; the production state interpretation is preserved.

## V.3 Conclusion

BLGC separates retained memory capacity from the work of a specified transition: $\ell^\infty$ accommodates uniformly bounded local state, projected maps preserve admissibility, and common-realization Lipschitz estimates control finite-horizon sensitivity. Bounded acquisition and evaluation yield the transition-work contract.

OPAL demonstrates tested temporal accounting, replay, preservation, isolation and recovery. Six million-node workloads passed, followed by controlled production-transition measurements at 10,000, 100,000 and one million nodes.

The receipt-index repair preserves byte-identical canonical state and recovery continuation while removing repeated full-index reconstruction. All 18 matched processes pass their correctness gates. Resident-update medians remain nearly unchanged; repaired receipt medians span 4.458-6.459 microseconds. Separate tests quantify capacity growth.

The contribution joins a mathematical foundation to a working engine with measured scaling and demonstrated operational reliability. Carried state, ordered continuation and incremental acquisition allow growing retained memory without forcing every ordinary event to revisit the collection.

## Reference scope

These seven works retain the original paper's intellectual background. They are not citations to the private production engine, nor are they evidence that its complete state satisfies the present mathematical hypotheses. Empirical tables derive from the recorded OPAL evidence described in Section IV.

The use of annotated references preserves the original explanatory purpose while separating background literature from the proofs and measurements actually supplied by this manuscript.

# Appendix A. Original benchmark figure

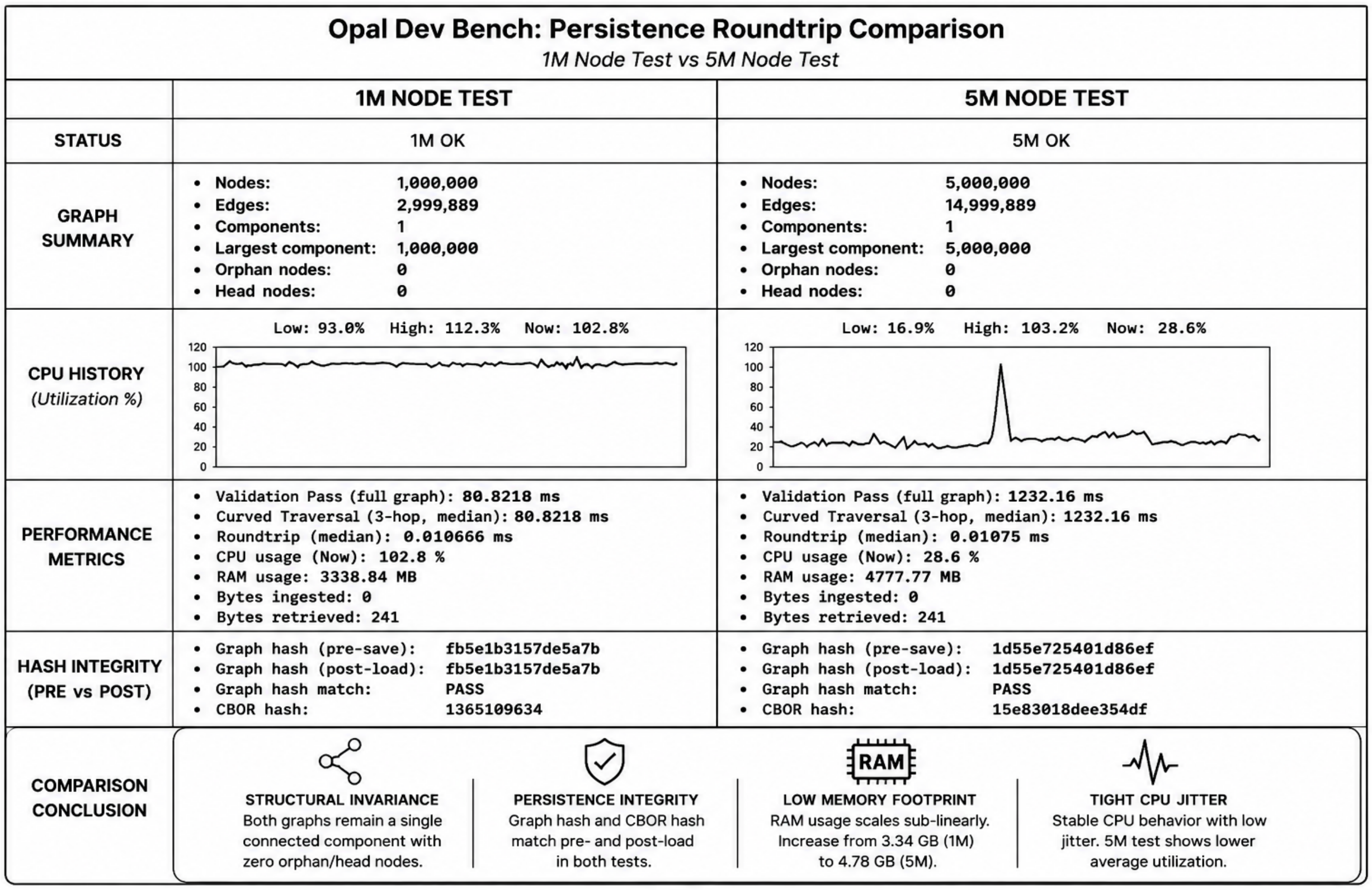

FIG 1

**Figure A1.** Original V.2 benchmark image, reproduced as a historical artifact. Its embedded labels and conclusions belong to that earlier presentation and are not adopted as current measurement claims. In particular, it displays 14,999,889 edges at five million nodes, unlike the later historical report summarized in IV.6. Its timing labels and RAM display must not be substituted for the definitions or measurements in Tables 1-9.

The image is retained to preserve the original record and visual explanation. It is not a matched before/after comparison, a temporal-evolution benchmark, or evidence of proportional bandwidth scaling. The current six-run access figure appears separately as Figure 1.

# Appendix B. Continuity of the revised construction

This revision preserves the original progression from local state to local generators, ordered evolution, state preservation, bounded work, and implementation evidence. It also retains the original bounded-increment observation, the identity convention g(0) = Id, the asynchronous execution explanation, the hardware motivation, the original figure, and all seven background references.

The main mathematical replacement is the supplied Banach-space formulation. Uniformly bounded configurations are represented in $\ell^\infty$ rather than presumed square-summable. Nonlinear projected generators are analyzed as Lipschitz coordinate maps rather than bounded linear or rank-one operators. Finite dimension dM is stated only for finite instances.

Temporal locality is the governing interpretation. A graph radius is an optional specialization, not its definition. Fixed active support is an explicit theorem hypothesis, and context acquisition is accounted for separately. Uniform functional-evaluation cost is assumed explicitly rather than inferred from regularity.

The finite-horizon results do not claim asymptotic contraction. State-dependent living-topology selection is treated as discrete switching unless continuity of the complete selector is proved. Numerical error and physical latency require additional, clearly separated conditions.

Historical metric corrections remain beside their datasets: median and P50 are separate, memory observations retain their sampling definitions, replay claims retain their tested scope, and bandwidth extrapolation remains distinct from measurement. IV.1 and IV.18 collect the correspondence and research boundaries.

This evidence revision preserves the supplied V.3 PDF, the full redraft, and the preceding critique revision as separate source artifacts. It retains the formal construction and earlier results while adding the completed production-transition study and its positive and negative findings.